\documentstyle[aps,preprint]{revtex}
\begin{document}
\draft
\def\bsigma{\mbox{\boldmath$\sigma$}}
\def\half{\mbox{$\frac{1}{2}$}}
\def\hd{\hat{\Delta}}
\def\hg{\hat{\gamma}}
\def\htrans{\hat{T}}

\preprint{cond-mat/9411065}
\title{Integrals of motion of the Haldane Shastry Model}
\author{J.C. Talstra and F.D.M. Haldane}
\address{Joseph Henry Laboratories, Princeton University, Princeton,
NJ 08544  USA}
\date{\today}

\maketitle
\begin{abstract}
In this letter we develop a method to construct all the integrals of
motion of
the $SU(p)$ Haldane-Shastry model of spins, equally spaced around a
circle,
interacting through a $1/r^2$ exchange interaction. These integrals
of motion
respect the Yangian symmetry algebra of the Hamiltonian.
\end{abstract}

The past few years have seen an extensive study of exactly solvable
quantum
many body systems with $1/r^2$-interactions. The simplest member of
this family
is the $SU(p)$ Haldane-Shastry model (HSM) with Hamiltonian
\cite{H88,S88}:
\begin{equation}
H_2 = -\sum_{i\neq j} \frac{z_i z_j}{(z_i -z_j)^2} P_{ij}
\end{equation}
describing $N$ particles with an internal spin degree of freedom that
can take on
$p$ different values, residing on equally spaced sites on a ring:
$\{ z_j =
\exp \left( \frac{2\pi i}{N} j\right)\}$.  For $p=2$ it describes
spin
$\frac{1}{2}$ particles.  The operator $P_{ij}$ permutes the spin of
two
particles at sites $i$ and $j$.  The energy levels of this model turn
out to
have huge degeneracies (beyond the regular global $SU(2)$ symmetry),
signaling
the presence of a large non-trivial symmetry algebra.  In Ref.\
\cite{HHTBP92} this
algebra was identified as the {\em Yangian}, a Hopf algebra
introduced by
Drinfel'd in 1986 \cite{Dr86}.  It describes the elementary
excitations of
this model:  {\em spinons}.  These spinons obey semion fractional
statistics.

The fact that this Yangian algebra commutes with the Hamiltonian
hints at the
integrability of this model.  However, the traditional method of
proving
integrability, i.e.\ construction of a set of commuting extensive
hermitean
operators $\{H_1,H_2,\ldots\}$, so-called invariants, has so far been
unsuccessful.  Invariants up to $H_4$ have been `guessed'
\cite{Ino90,HHTBP92}.
Minahan and Fowler \cite{MF93} introduced a set of invariants that
commute
with the Hamiltonian, based on operators introduced by Polychronakos
\cite{Poly92}.  However, the generating function for this set is
essentially
the trace of the transfer matrix and thus contains only elements of
the
Yangian algebra.\footnote{The authors of Ref.\protect\cite{MF93}
claim to have
found the Hamiltonian $H_2$ in their third order invariant, but in
reproducing
their algebra we did not find any such term; in fact we only
recovered yangian
operators.}

In this letter we will construct a set $\{ H_n\}$ of extensive
operators that
commute among themselves and with the Yangian.  In order to do this
we have to
consider a more general {\em dynamical} model in which the particles
are
allowed to move along the ring:  the Calogero Sutherland model (CSM)
with an
internal degree of freedom.  This model, which has been studied in
\cite{Poly92,HH92} has the following Hamiltonian:

\begin{equation}
H=\sum_{j=1}^{N} \hbar^2 \left(z_j \frac{\partial}{\partial
z_j}\right)^{2} -
\sum_{i\neq j} \lambda (\lambda +P_{ij} ) \frac{z_i z_j}{(z_i -z_j
)^2}.
\label{dynmodeldef}
\end{equation}
When $\lambda\rightarrow\infty$ or equivalently $\hbar\rightarrow 0$,
the
particles `freeze' into their classical equilibrium positions, and
barring some
subtleties we recover the spin Hamiltonian $H_2$.

The reason for this diversion through the dynamical model to obtain
the
constants of motion is the following:  the so-called {\em quantum
determinant}
of the transfer matrix, an object that commutes with the Yangian
algebra and
therefore a natural candidate for the generating function of the
constants of
motion, happens to be {\em scalar} in the spin model (i.e.\ when
$\hbar\rightarrow 0$), as we shall see below.  But in the general
dynamical
model, this is not the case, and by carefully taking the limit
$\lambda\rightarrow\infty$ we can isolate a generating function for
the $\{
H_n\}$.

Let us first review the r\^{o}le of the Yangian algebra in the
dynamical model.
A more extensive treatment can be found in \cite{H94,BGHP93}.  The
integrability of the CSM is based on the existence of the {\em
transfer-matrix}
$T^{ab}(u)$ that commutes with the Hamiltonian:

\begin{eqnarray}
T^{ab}(u) &=& \delta^{ab} + \sum_{n=0}^{\infty}
\frac{\lambda}{u^{n+1}}
T^{ab}_n \nonumber\\ T^{ab}_{n} &=& \sum_{i,j=1}^{N}X^{ab}_i (L^n
)_{ij}\nonumber\\ L_{ij}
&=&
\delta_{ij}z_j\partial_{z_j}+(1-\delta_{ij})\lambda\theta_{ij}P_{ij},
\end{eqnarray}
where $X^{ab}_j$, $a,b=1,\ldots ,p$ acts as $|a\rangle\langle b|$ on
the spin
of particle $i$, and $\theta_{ij}=\frac{z_i}{z_i-z_j}$.
This transfer matrix satisfies the Yang-Baxter equation:

\begin{equation}
R_{00'}(u-v) T^{0}(u) T^{0'}(v) = T^{0'}(v) T^{0}(u) R_{00'}(u-v),
\label{YBdef}
\end{equation}
with $R_{00'}(u)=u+\lambda P_{00'}$ and $T^0 (u)=T(u)\otimes 1$ and
$T^{0'}(u)=1\otimes T(u)$.  For the purposes of this letter we will
deal with
another form of the same transfer matrix.  Introduce the following
representation of the so-called {\em Dunkl}-operators \cite{BGHP93}:

\begin{eqnarray}
\hat{D}_i &\equiv&\hbar z_i\partial_{z_i} + \hat{\gamma}_i = \hbar
z_i\partial_{z_i} + \frac{\lambda}{2}\sum_{j(\neq i)}
(w_{ij}-{\rm sgn}(i-j))K_{ij} \nonumber\\
w_{ij} & = & \frac{z_i+z_j}{z_i-z_j}
\end{eqnarray}
$K_{ij}$ is the operator that permutes the {\em spatial} co-ordinates
of
particles
$i$ and $j$. These Dunkl-operators commute:

\begin{equation}
[\hat{D}_i ,\hat{D}_j ] = 0,
\end{equation}
but are not covariant under permutations:

\begin{eqnarray}
&& [K_{i,i+1},\hat{D}_k] = 0 \;\;\;{\rm if}\; k\neq i,i+1\nonumber\\
&& K_{i,i+1}\hat{D}_i - \hat{D}_{i+1} K_{i,i+1} = \lambda,
\label{dunklalg}
\end{eqnarray}
defining a so-called {\em degenerate affine Hecke Algebra}.
In terms of these Dunkl operators we can define a transfer matrix
that also
obeys the Yang-Baxter equation:

\begin{equation}
\hat{T}^0 (u) = \left(1+\frac{\lambda
P_{01}}{u-\hat{D}_1}\right)\cdots
\left(1+\frac{\lambda P_{0N}}{u-\hat{D}_N}\right).
\end{equation}
It satisfies eq.\ (\ref{YBdef}) trivially since the $\{ \hat{D}_i \}$
commute
amongst themselves and commute with the $P_{0j}$, since the latter
only act on
spin degrees of freedom; furthermore
$\left(1+\frac{\lambda P_{0i}}{u-\hat{D}_i}\right)$ is the elementary
transfer
matrix with spectral parameter $u-\hat{D}_i$.  To retrieve $T^0 (u)$
from
$\hat{T}^0 (u)$ we have to apply a projection $\Pi$ to $\hat{T}^0$
that
replaces every occurrence of $K_{i,i+1}$ by $P_{i,i+1}$ once ordered
to the
right of an expression (this is equivalent to having the unprojected
operator
act on wavefunctions that are symmetric under simultaneous
permutations of
spin- and spatial co-ordinates) \cite{BGHP93}. We will drop the ``0''
subscript
on $\hat{T}(u)$ from here on.

Normally the conserved quantities are derived from a Taylor expansion
of the
trace of the transfer matrix. In this case  that just gives us
combinations of
Yangian operators, elements of the symmetry algebra. This set doesn't
even contain the Hamiltonian. As pointed by various authors
\cite{Kor81,Dr86,BGHP93}, there is another quantity that commutes
with the
Hamiltonian, derivable from the transfer matrix: the quantum
determinant,
\begin{equation}
{\rm Det}_q (T(u)) = \sum_{\sigma\in S_p}\epsilon(\sigma )
T_{1\sigma_1}
(u-\lambda (p-1))T_{2\sigma_2}(u-\lambda (p-2))\cdots
T_{p\sigma_p}(u).
\end{equation}
It satisfies $[T(u),{\rm Det}_q(T(u)) ]=0$.  It has been computed in
\cite{BGHP93} as:

\begin{equation}
{\rm Det}_q (T(u)) = \Pi\: {\rm Det}_q (\hat{T}(u))\Pi = \Pi\:\left(
\frac{\hat{\Delta}(u+\lambda,\hbar)}{\hat{\Delta
(u,\hbar)}}\right)\Pi.
\end{equation}
where (making the dependence on $\hbar$ explicit):

\begin{equation}
\hd (u,\hbar )=\prod_{i} (u-\hat{D}_i(\hbar )).
\label{deltadef}
\end{equation}
Now obviously:

\begin{equation}
[\hd (u,\hbar),\hat{T} (v,\hbar) ] = 0.
\label{dtcommut}
\end{equation}
This holds since the $\hat{D}_i$'s commute with each other and the
$P_{0j}$'s .
The projector doesn't get in the way since a product of projections
is the
projection of the product---both $\htrans(v)$ and $\hd(u)$ are
symmetric under
simultaneous permutation of spin- and spatial co-ordinates
\cite{BGHP93}.  The
eigenvalues of $\hd(u)$ are also known:  for every partition $|n|$
there is an
eigenvalue:

\begin{equation}
\delta^{|n|}(u) = \prod_{j=1}^{N}(u-\hbar n_j -\lambda
(j-\frac{N+1}{2}))
\label{deltaev}
\end{equation}
We notice that as $\hbar\rightarrow 0$, i.e.\ in the limit of the
HSM, all
eigenvalues become identical and $\hd(u,0)$ is a {\em multiple of the
identity operator}.  Thus no
non-trivial constants of motion are contained in $\hd(u,0)$.
Nevertheless let
us study (\ref{dtcommut}) for small $\hbar$.  Writing
$\htrans(v,\hbar)=\sum_{n}\hbar^n \htrans_n (v)$;
$\hd(u,\hbar)=\sum_{n}\hbar^n
\hd_n(u)$:

\begin{eqnarray}
0 &=& [\htrans(v,\hbar),\hd(u,\hbar)]  \nonumber\\
&=&[\htrans_0(v),\hd_0(u)]+\hbar\left([\htrans_0(v),\hd_1(u)]
+[\htrans_1(v),\hd_0(u)]\right) +
{\cal O}(\hbar^2).
\label{hbarexp}
\end{eqnarray}
The ${\cal O}(\hbar ^0)$ term is trivially 0.  The rest of this
letter will
focus on the vanishing of the ${\cal O}(\hbar)$ term. As we shall
show below:
$[\htrans_1(v),\hd_0(u)]=0$.  Therefore we have the important result:
$[\htrans_0(v),\hd_1(u)]=0$, i.e.\ the ${\cal O}(\hbar)$ term in
$\hd(u,\hbar)$
commutes with the transfer matrix (and therefore the Yangian) of the
Haldane-Shastry spin model as well.  Furthermore it will also become
apparent that

\begin{equation}
[ \hd_1(u),\hd_1(u') ] =0 .
\end{equation}
So we can take $\hd_1(u)$ to be the generating function of the
constants of
motion of the HSM!  In order to establish these results we first need
to prove
the following corollary:  $z_i\partial_{z_i} \hd_0(u)=0$ when
evaluated with the particles at their equilibrium positions, i.e.\
$z_j=\exp \left( \frac{2\pi i}{N} j\right)$.

{}From eq.\ (\ref{deltadef}) we have $\hd_0(u)=\prod_{i} (u-\hg_i)$.
Since we
know that $\hd_0(u)$ is scalar we can evaluate it by having it act on
any
convenient state, e.g.\ the one where all particles have identical
values for
their internal degree of freedom (for $p=2$ we would say:  all spins
pointing
up).  I.e.\ the permutations reduce to 1.  This has been done in
\cite{BGHP93}:

\begin{eqnarray}
\hd_0(u)& = & {\rm det}(u- \Theta)\nonumber\\
\Theta_{ij} & = & \frac{\lambda z_i}{z_i -z_j} (1-\delta_{ij}).
\end{eqnarray}
Then using $\partial_x {\rm det}A(x) = {\rm Tr}[A^{-1}\partial_x
A(x)] {\rm
det}[A(x)]$ we have:

\begin{eqnarray}
-\partial_{z_i}\hd_0(u) &=&
{\rm Tr}\left[\frac{1}{u-\Theta}\partial_{z_i}\Theta\right]{\rm det}
[u-\Theta]
\nonumber\\
&=& \sum_{n=0}^{\infty} u^{-(n+1)}
{\rm Tr}\left[\Theta^n\partial_{z_i}\Theta\right] {\rm det}[u-\Theta
].
\end{eqnarray}
Now evaluate the trace in a basis where $\Theta$ is diagonal. This
can clearly
be done for $\hbar\rightarrow 0$ and $z_j = \exp \left( \frac{2\pi
i}{N}
j\right)$. $\Theta$ has eigenvectors $\psi_n$ , where $(\psi_n)_j =
\frac{1}{\sqrt{N}}\exp \left(\frac{2\pi i}{N} jn\right)$, with
eigenvalue
$\lambda\left(\frac{N+1}{2}-n\right)$. Then

\begin{equation}
{\rm Tr}[\Theta^m \partial_{z_i}\Theta ] =
\sum_{n=1}^{N}\langle\psi_n|\partial_{z_i}\Theta|\psi_n\rangle
\lambda^m \left(\frac{N+1}{2}-n\right)^m.
\end{equation}
Using that
\begin{equation}
(\partial_{z_i}\Theta )_{jk} =  \left\{
\begin{array}{ll}
\frac{\lambda}{(z_i -z_j)^2}(z_j\delta_{ki}-z_k\delta_{ij}) & j\neq k
\\
0 & \mbox{otherwise}
\end{array}
\right.
\end{equation}
we find:

\begin{eqnarray}
\langle\psi_n |\partial_{z_i}\Theta |\psi_n\rangle &=&
\sum_{p(\neq i )}\frac{\lambda}{N}
\frac{e^{\frac{2\pi i}{N}(i-p)n} - e^{\frac{2\pi i}{N}(p-i)n}}{\left(
e^{\frac{2\pi i}{N} p}-e^{\frac{2\pi i}{N} i}\right)^2} e^{\frac{2\pi
i}{N} p}\nonumber\\
&=&\frac{i\lambda}{2N}\sum_{p=1}^{N-1}\frac{\sin\left(\frac{2\pi
pn}{N}\right)}{\sin^2\left( \frac{\pi p}{N}\right)} \equiv 0.
\end{eqnarray}
Therefore $\partial_{z_i}\hd(u)_0 = 0$ at $z_j=\exp \left( \frac{2\pi
i}{N}j\right)$.

{}From expanding $\hd(u,\hbar)$ to ${\cal O}(\hbar)$ in
(\ref{deltadef}) we have

\begin{equation}
 \hd_1(u) = \sum_{i=1}^{N}\left(\prod_{j=1}^{i-1}(u-\hg_j)\right)
\: z_i\partial_{z_i} \: \left(\prod_{j=i+1}^{N}(u-\hg_j)\right).
\end{equation}
Then, with the corollary and the fact that $[\hg_i ,\hg_j ]=0$ for
all $i,j$
(the Dunkl algebra (\ref{dunklalg}) is satisfied for $\hbar =0$ as
well):

\begin{eqnarray}
[\hd_1(u),\hd_0(u)]&=&\sum_{i=1}^{N}\left(\prod_{j=1}^{i-1}
(u-\hg_j)\right)
[z_i\partial_{z_i},\hd_{0}(u)]\left(\prod_{j=i+1}^{N}(u-\hg_j)\right)
\nonumber\\
&=&\sum_{i=1}^{N}\left(\prod_{j=1}^{i-1}(u-\hg_j)\right)
\left\{ z_i\partial_{z_i} \hd_0(u)\right\} \left(\prod_{j=i+1}^{N}
(u-\hg_i)\right) = 0
\label{delzeroone}
\end{eqnarray}
With this result we find:

\begin{eqnarray}
\frac{1}{\hd(u,\hbar)}&=&\frac{1}{\hd_0(u)}-\hbar
\frac{1}{\hd_0(u)}\hd_1(u)
\frac{1}{\hd_0(u)} + {\cal O}(\hbar^2) \nonumber\\
&=& \frac{1}{\hd_0(u)}-\hbar
\frac{1}{\left(\hd_0(u)\right)^2}\hd_1(u) +
{\cal O}(\hbar^2).
\end{eqnarray}
This can be checked by multiplying the LHS and RHS by $\hd(u,\hbar)$.
Now let
us expand $\htrans(v,\hbar)$ to ${\cal O}(\hbar)$:

\begin{eqnarray}
\htrans(v,\hbar)&=& \prod_{i=1}^{N} \left( 1+\frac{\lambda
P_{0i}}{v-\hat{D}_i}
\right) =
\frac{1}{\hd(v,\hbar)}\prod_{i=1}^{N} (v-\hat{D}_i +\lambda P_{0i})
= \frac{1}{\hd_0(v)}\prod_{i=1}^{N}(u-\hg_i +\lambda P_{0i}) +
\nonumber\\
&& +
\hbar\left\{-\left(\frac{1}{\hd_0(v)}\right)^2
\hd_1(v)\prod_{i=1}^{N}\left( v-\hg_i
+\lambda P_{0i}\right)\right. \nonumber\\
&& \left. -\frac{1}{\hd_0(v)}
\sum_{i=1}^{N}\left(\prod_{j=1}^{i-1}(v-\hg_j +\lambda P_{0j})\right)
z_i\partial_{z_i} \left(\prod_{j=i+1}^{N}(v-\hg_j +\lambda
P_{0j})\right)
\right\}+{\cal O}(\hbar^2)\nonumber\\
&\equiv& \htrans_0(v)+\hbar\htrans_1(v) + {\cal O}(\hbar^2)
\label{trmexpand}
\end{eqnarray}
It is now obvious how $[\htrans_1(v),\hd_0(u)]=0$.  The contribution
from the
first term in curly brackets in (\ref{trmexpand}) vanishes by virtue
of eq.\
(\ref{delzeroone}).  As far as the second term is concerned:  the
$\hg_i$
commute amongst each other and with the $P_{0j}$, and
$[z_i\partial_{z_i},\hd_0(u)]=0$ as we showed before.

So far we have established that $\hd_1(u)$ respects the Yangian
symmetry, but
we also need to show that it is a {\em good} generator of constants
of motion
in that it commutes with itself at a different value of the parameter
$u$:
$[\hd_1(u),\hd_1(u')]=0$.  It will be enough to prove this on the
space spanned
by the Yangian Highest Weight states (YHWS).  These states are, as
their name
implies, the highest weight states of a representation of the Yangian
algebra.
All other states in the model are generated by acting on these YHWS
with the
elements of the Yangian algebra, i.e.\ the transfer matrix.  Since
$\hd_1(u)$
commutes with $\htrans_0(v)$ for any $u$, $[\hd_1(u),\hd_1(u')]=0$
will
therefore also hold on the non-highest weight states.  First of all
we should
note that $\hd_1(u)$ and $\hd_1(u')$ leave the space of YHWS
invariant.  This
follows from the fact that all such states $|\Gamma\rangle$ are
annihilated by
$\htrans_0^{ab}(v)$ with $a>b$ (see ref.  \cite{BGHP93}).  But since
$\htrans_0^{ab}(v)$ commutes with $\hd_1(u)$,
$\htrans_0^{ab}\left(\hd_1(u)|\Gamma\rangle\right)$ will also be 0
for $a>b$.

The proof that $\hd_1(u)$and $\hd_1(u')$ commute hinges on the
existence of an
operator that commutes with both these $\hd_1$'s and is {\em
non-degenerate}.  Such an operator is $T^{pp}_0(v)$.  Its eigenvalues
are given
by \cite{BGHP93}:

\begin{equation}
\frac{ P_1(\overline{v}+1)\cdots P_{p-1}(\overline{v}+1)}{
P_1(\overline{v})\cdots P_{p-1}(\overline{v})}\;\;\;\; {\rm with}\;
\overline{v}=v/\lambda
\label{evform}
\end{equation}
The polynomials $P_{i}(\overline{v}),\; i=1\ldots p-1$ characterize
the
representation of the Yangian.  As was found in \cite{BGHP93}, every
degenerate
supermultiplet in the $SU(p)$ HSM (i.e.\ every representation of the
Yangian)
can be represented by a string of $N+1$ binary digits 0 or 1, called
a {\em
motif}.  The first and last (entry 0 and $N$) are always 0.  For an
$SU(p)$
model the string cannot contain more than $p-1$ consecutive ones.
A set of $k-1$
consecutive 1's is called a $k$-string.  To generate the polynomials
$P_i$
replace very 0 in the motif by `)('.  We then have groups of 1's
enclosed in
parentheses.  The polynomial $P_k(\overline{v}$) has zeroes at
$\frac{1}{2}$ +
rightmost member of a $k$-string.  Let us consider an example to
clarify this:
for $SU(3)$, $N=14$ the motif 011010001100100 has four 1-strings, two
2-strings
and two 3-strings.  This implies polynomials
%% FOLLOWING LINE CANNOT BE BROKEN BEFORE 70 CHAR
$P_1(\overline{v})=(\overline{v}-5\half)(\overline{v}-6\half)(\overline{v}-
10\half)(\overline{v}-13\half)$,
$P_3(\overline{v})=(\overline{v}-4\half)(\overline{v}-12\half)$ and
$P_3(\overline{v})=(\overline{v}-2\half)(\overline{v}-9\half)$.

The eigenvalues (\ref{evform}) are obviously independent rational
functions of
$\overline{v}$, and $T_0^{pp}(v)$ is non-degenerate.  If
$|\Gamma\rangle$ is a
YHWS with motif $\Gamma$ then $\hd_1(u)|\Gamma\rangle$ and
$\hd_1(u')|\Gamma\rangle$ both are scalar multiples of
$|\Gamma\rangle$
since they have the same $T^{pp}_0(v)$-eigenvalue
($[T^{pp}_0(v),\hd_1(u)]=
[T^{pp}_0(v),\hd_1(u')]=0$).  So in this
YHWS-space both $\hd_1(u)$ and $\hd_1(u')$ are diagonal and thus
commute.

In the remaining part we will reproduce the integrals of motion that
have
already been found \cite{Ino90,HHTBP92}, and point out some
subtleties in their
construction.  As is customary for the Heisenberg model with nearest
neighbor
exchange we take $\Gamma_{1}(u)=\frac{d}{du} \ln(\hd_1(u) )$ rather
than
$\hd_1(u)$ to be the generating function for the integrals of motion,
so
that the invariants will have an additive spectrum.  When expanded in
powers of $u$ it
reads:

\begin{eqnarray}
\Gamma_1(u) &=& \Pi \sum_{i=1}^{N}\frac{1}{u-\hbar z_i\partial_{z_i}
-\hg_i}
\Pi\;\; {\rm  to}\;{\cal O}(\hbar) \nonumber\\
 & = & \sum_{n=0}^{\infty}u^{-(n+1)}\left\{
\Pi\sum_{i=1}^{N}\sum_{p=0}^{n-1}
(\hg_i)^p z_i\partial_{z_i} (\hg_i)^{n-p-1} \Pi\right\}\nonumber\\
&\equiv &\sum_{n=0}^{\infty}u^{-(n+1)} H_n,
\label{Hgendef}
\end{eqnarray}
where we have reinserted the projection operator that turns
$K_{ij}\rightarrow
P_{ij}$, when ordered to the right of an expression.
We have worked out the first few $H$'s. With $z_{ij} = z_i -z_j$ we
have:

\begin{eqnarray}
H_1 &=& \sum_{i=1}^{N} z_i\partial_{z_i} = P\nonumber\\
H_2 &=& -\sum_{i,j}\rule{0in}{1.3ex}^{'}\,
\frac{z_i z_j}{z_{ij}^2} (P_{ij}-1)\nonumber\\
H_3 &=& \sum_{ijk}\rule{0in}{1.3ex}^{'}\,
\frac{z_i z_j z_k}{z_{ij}z_{jk}z_{ki}} P_{ijk} +
\frac{3}{4} \sum_{ij}\rule{0in}{1.3ex}^{'}\,
(1-(w_{ij})^2)z_i\partial_{z_i} = \nonumber\\
&=& \sum_{ijk}\rule{0in}{1.3ex}^{'}\,
\frac{z_i z_j z_k}{z_{ij}z_{jk}z_{ki}} P_{ijk} + \frac{N^2
-1}{4} P \nonumber\\
H_4&=& \sum_{ijkl}\rule{0in}{1.3ex}^{'}\,
\frac{z_i z_j z_k z_l}{z_{ij}z_{jk}z_{kl}z_{li}}(P_{ijkl}-1)
-2 \sum_{ij}\rule{0in}{1.3ex}^{'}\,
\left(\frac{z_i z_j}{z_{ij} z_{ji}}\right)^2 (P_{ij}-1)
+\frac{N^2-1}{3} H_2
\label{Hexpl}
\end{eqnarray}
The prime on the summation symbol indicates that the sum should be
restricted
to distinct summation-indices.  To compute the previous expressions
we normal
ordered the $z_i \partial_{z_i}$ to the right in eq.\ (\ref{Hgendef})
and {\em
then} put $z_j = \exp\left(\frac{2\pi i}{N} j\right)$.  The identity
$w_{ij}w_{jk}+w_{jk}w_{ki}+w_{ki}w_{ij}=-1$ which lies at the heart
of the
integrability of these $\frac{1}{r^2}$-models is very useful in the
reduction
of these expressions.  $P$ indicates the momentum and we will show
its
interpretation later on.  Notice the absence of Yangian operators as
well as
terms containing both permutations {\em and} derivatives.
The expressions in eq.\
(\ref{Hexpl}) can be seen to coincide with those reported previously
\cite{HHTBP92},\footnote{We should point out a correction in eq.\ (7)
of
Ref.\ \protect\cite{HHTBP92} where $-\frac{1}{3} H_2$ should be
replaced by
$+\frac{1}{6} H_2$.  This changes the invariant in a harmless manner,
but this is relevant for
comparing the eigenvalues of the operators in that article and the
ones
that we will find later
on.}  lending credibility to this way of deriving the integrals of
motion.
Unfortunately for large $n$ it becomes prohibitively complicated to
compute
$H_n$.

We have an alternative way to verify the validity of these constants
of motion
as well. We will proceed to compute the eigenvalues of the operators
$H_n$  and
compare these to the `rapidity' description of the eigenvalues in
\cite{HHTBP92}. We will constrain ourselves to the $SU(2)$ case to
simplify the
algebra.

As is well known \cite{KR86} the roots of $\hd(u,\hbar)$---i.e.\ the
poles of
the transfer matrix $\htrans(u,\hbar)$, see eq.\
(\ref{trmexpand})---are given
by the solutions of so-called Bethe Ansatz equations, which only
depend on the
{\em two}-particle phase-shift.  In the case of the CSM it is
$\pi\lambda {\rm
sgn}(k_1 -k_2)$.  Notice that this phase-shift only depends on the
ordering of
the momenta $k_1$ and $k_2$.  This is why these models are
interpreted to
describe an ideal gas of particles with statistics that interpolates
between
bosons ($\lambda =0$) and fermions ($\lambda = 1$).  In the
dynamical model (\ref{dynmodeldef}) the particles have charge {\em
and} spin.
Therefore we get two coupled sets of `Nested' Bethe Ansatz
equations---for the general case $p\neq 2$ there are $p$ equations.
They
have been presented in \cite{Kaw93}:

\begin{eqnarray}
k_i L& =&\lambda \left\{\sum_{j(\neq i)}\pi{\rm sgn}(k_i -k_j) +
\frac{1}{\lambda}\left[ 2\pi I_i - \pi\sum_{\alpha =1}^{M}
{\rm sgn}(k_i -\Lambda_{\alpha})\right]\right\}\nonumber\\
&\equiv & \left( k_i^0 +\frac{1}{\lambda}\delta k_i\right) L
\label{kdef}\\
&&\hspace{-.75truein}
\pi\sum_{\beta(\neq\alpha)}{\rm
sgn}(\Lambda_{\alpha}-\Lambda_{\beta}) +
2\pi J_{\alpha} = \pi\sum_{i=1}^{N}{\rm sgn}(\Lambda_{\alpha}-k_i).
\label{lamdef}
\end{eqnarray}
We have reinstated $L$, the circumference of the circle to get the
dimensions
correctly.  Notice how $\hbar$ drops out of these equations due to
the fact
that the full Hamiltonian is scale invariant (a peculiarity of the
$\frac{1}{r^2}$-type potentials).  So rather than sending
$\hbar\rightarrow 0$
we should let $\lambda\rightarrow \infty$.  There are $N$ equations
defining
the $\{ k_i\}$ (one for every particle) with charge quantum numbers
$\{ I_i\}$.
Furthermore we have $M$ equations defining the auxiliary momenta $\{
\Lambda_{\alpha}\}$.  $M$ is the number of particles with a spin
$\downarrow$.
The $\{J_{\alpha}\}$ are their {\em spin} quantum numbers.  The I's
and J's are
distinct integers or half-odd integers depending on the parity of $N$
and
$M-N$.

Furthermore we should restrict the Hilbert space to states that only
carry spin
excitations, and no charge excitations (these elastic modes
unfortunately don't
acquire a gap as $\hbar\rightarrow 0$)\footnote{This is why we don't
use eq.\
(\protect\ref{deltaev}) which gives a much more direct expression for
the
eigenvalues of
$\hd(u,\hbar)$; however, one doesn't know a priori whether an
eigenvalue
belongs to a pure
spin- or charge excitation, or a mixture of both.}.  As in for
instance the
1D-Hubbard model we accomplish this by leaving the charge quantum
numbers in
their groundstate configuration and only exciting the $\{
J_{\alpha}\}$.  Let
us therefore first analyze the absolute groundstate which has
$M=\frac{N}{2}$
(for $N$ even), so there are twice as many $k$'s as $\Lambda$'s.  The
$I$'s and
$J$'s are consecutive and spaced by one unit.  Then eq.\
(\ref{lamdef}) tells
us that between every two $\Lambda$'s there must be two $k$'s.  In a
spin
excited state we have $M<\frac{N}{2}$ and by leaving openings in the
$J$-distribution, we can have more than two $k$'s sit between every
pair of
$\Lambda$'s.  Notice that this equation doesn't fix the value of the
$\Lambda$'s, just their positions with respect to the $k_i$'s.  From
eq.\
(\ref{kdef}) we learn that when we order the $\{ k_i\}$ such that
$k_i < k_j$
for $i<j$:  $k_i \approx \lambda \left( i-\frac{N+1}{2}\right)\equiv
k_i^0$.
There is however an ${\cal O}(\frac{1}{\lambda})={\cal O}(\hbar)$
correction
through the $\Lambda$'s.  Whenever a $\Lambda$ sits between two $k$'s
they will
be drawn together by $\frac{1}{\lambda}$.  This information is
contained in
$\delta k_i$.  Now, in the same way that the constants of motion are
contained
in the ${\cal O}(\hbar)$-term in $\hd(u,\hbar)$, their eigenvalues
are
determined by the ${\cal O}(\frac{1}{\lambda})$ corrections to the
$k$'s.  As
$\hd(u,\hbar)$ has eigenvalues $\prod_{i=1}^{N}(u-k_i)$,
$\Gamma(u,\hbar)=\frac{d}{du} \ln\hd(u)\equiv\Gamma_0
+\hbar\Gamma_1+\ldots$
must
have eigenvalues $\sum_{i} \frac{1}{u-k_i}=\sum_i \frac{1}{u-k_i^o -
\frac{1}{\lambda}\delta k_i}$.  So for $\Gamma_1(u)$ acting on a
state
characterized by a set $\{ \Lambda_{\alpha}\}$ we find its
eigenvalue:

\begin{eqnarray}
\Gamma_1(u) &=& \sum_{n=0}^{\infty}
\frac{1}{u^{n+1}} n \sum_{i=1}^{N}(k_i^0)^{n-1} \delta
k_i\{\Lambda_{\alpha}\}
\nonumber\\
&\equiv& \sum_{n=0}^{\infty} \frac{1}{u^{n+1}}
h_n\{\Lambda_{\alpha}\}.
\end{eqnarray}
We will now label the $\Lambda_{\alpha}$'s by $m_\alpha$, their
positions
relative to the $k$'s, i.e.\ if $\Lambda_\alpha$ sits between $k_r$
and
$k_{r+1}$, then $m_\alpha = r$. We see  that the $m_\alpha$ have to
be at least
two units apart since there are {\em at least} two $k$'s between
$\Lambda$'s
according to eq.\ (\ref{lamdef}). Now writing $\frac{1}{2}
{\rm sgn}(\Lambda_\alpha -k_i) = \theta(\Lambda_\alpha -k_i) -\half$
($\theta$
is the step function), we have:

\begin{eqnarray}
&& n\sum_{i=1}^{N}(k_i^0)^{n-1}\delta k_i\{\Lambda_\alpha\} =
n\sum_{i=1}^{N}
\left( i-\frac{N+1}{2}\right)^{n-1}\left\{ I_i-\frac{M}{2} +
\sum_{\alpha =1}^{M} \theta (\Lambda_\alpha -k_i)\right\}\nonumber\\
&&={\rm const} + \sum_{\alpha =1}^{M} \left[ n
\sum_{i=1}^{m_{\alpha}}
\left( i-\frac{N+1}{2}\right)^{n-1} \right] \nonumber\\
&&={\rm const}+\sum_{\alpha =1}^{M} \epsilon_m (m_\alpha)
\end{eqnarray}

For small $n$ we can evaluate $\epsilon_n(m_\alpha)$ exactly:

\begin{eqnarray}
\epsilon_1(m_\alpha) &=& m_\alpha\nonumber\\
\epsilon_2(m_\alpha) &=& m_\alpha (m_\alpha -N)\nonumber\\
\epsilon_3(m_\alpha) &=& \half m_\alpha (m_\alpha -N)(2m_\alpha -N) +
\frac{N^2-1}{4} m_\alpha\nonumber\\
\epsilon_4(m_\alpha) &=& \epsilon_2(m_\alpha) \left(
\epsilon_2(m_\alpha) +
\frac{N^2-1}{2}\right)
\end{eqnarray}

These results coincide nicely with the numerical values of
\cite{HHTBP92}, when
we interpret the  $m_\alpha$ as the rapidities of the HSM! We notice
that it is
indeed consistent to interpret the momentum term $P$ in (\ref{Hexpl})
in $H_1$
and $H_3$ as  $\sum_{\alpha} m_\alpha$, i.e.\ the degree of the
polynomial
wavefunction.

In conclusion we have outlined a method to obtain the constants of
motion of
the HSM as a strong coupling limit of the CSM with particles with
internal
degrees of freedom.  Although the task to actually obtain the
invariants is
quite cumbersome, it can be done in principle.  Given the relatively
simple
structure of the invariants we expect there to be some technique that
could
simplify the computation considerably.  The construction of integrals
of motion
presented in this letter, provides us with extensive operators that
commute
with each other and the Yangian symmetry algebra.  By computing
eigenvalues of
the invariants through the Nested Bethe Ansatz, and comparing them
with
previous numerical results \cite{HHTBP92} we provided evidence for
the validity
of the approach.  It would be interesting to analyze the cause of the
miraculous absence of terms containing mixtures of permutations and
derivatives.

This work was supported in part by NSF Grant No.\ DMR922407.

\end{document}